\documentstyle[amssymb,preprint,aps]{revtex}

\begin{document}
\draft
\title{An interferometric complementarity experiment in a bulk Nuclear Magnetic
Resonance ensemble}
\author{Xinhua Peng$^{1\thanks{%
Corresponding author. {E-mail:xhpeng@wipm.ac.cn; Fax: 0086-27-87885291.}}}$,
Xiwen Zhu$^{1}$, Ximing Fang$^{1,2}$, Mang Feng$^{1}$, Maili Liu$^{1},$ and
Kelin Gao$^{1}$}
\address{$^{1}$State Key Laboratory of Magnetic Resonance and Atomic and Molecular\\
Physics,\\
Wuhan Institute of Physics and Mathematics, The Chinese Academy of Sciences,%
\\
Wuhan, 430071, People's Republic of China\\
$^{2}$Department of Physics, Hunan Normal University, Changsha, 410081,\\
People's Republic of China}
\maketitle

\begin{abstract}
We have experimentally demonstrated the interferometric complementarity,
which relates the distinguishability $D$ quantifying the amount of which-way
(WW) information to the fringe visibility $V$ characterizing the wave
feature of a quantum entity, in a bulk ensemble by Nuclear Magnetic
Resonance (NMR) techniques. We primarily concern on the intermediate cases:
partial fringe visibility and incomplete WW information. We propose a
quantitative measure of $D$ by an alternative geometric strategy and
investigate the relation between $D$ and entanglement. By measuring $D$ and $%
V$ independently, it turns out that the duality relation $D^{2}+V^{2}=1$
holds for pure quantum states of the markers.
\end{abstract}

\pacs{PACS numbers: 03.65.Ud, 03.67.-a}

\vskip 1cm

\narrowtext

\section{\bf Introduction}

Bohr complementarity\cite{Bohr} expresses the fact that quantum systems
possess properties that are equally real but mutually exclusive. This is
often illustrated by means of Young's two-slit interference experiment,
where ``the observation of an interference pattern and the acquisition of
which-way (WW) information are mutually exclusive''\cite{Englert}. As stated
by Feynman, the two-slit experiment ``has in it the heart of quantum
mechanics. In reality it contains the only mystery''\cite{Feynman}.
Complementarity is often superficially identified with the `wave-particle
duality of matter'. As its tight association with the interference
experiment, the terms of the ``interferometric duality'' or
``interferometric complementarity'' are more preferable. Two extreme cases,
``full WW information and no fringes when measuring the population of
quantum states'' and ``perfect fringe visibility and no WW information''
have been clarified in textbooks and demonstrated with many different kinds
of quantum objects including photons\cite{Taylor}, electrons\cite
{Mollenstedt}, neutrons\cite{Zeilinger}, atoms\cite{Carnal} and nuclear
spins in a bulk ensemble with NMR techniques\cite{Zhu}. In Ref.[8] we
further proved theoretically and experimentally that full WW information is
exclusive with population fringes but compatible with coherence patterns.

In order to describe the duality in the intermediate regime ``partial fringe
visibility and partial WW information'', quantitative measures for both the
fringe visibility $V$ and WW information are required. The definition of the
former is the usual one. In variants of two-slit experiments different WW
detectors or markers, such as microscopic slit and micromaser, are used to
label the way along which the quantum entity evolves. A quantitative
approach to WW knowledge was first given by Wootters and Zurek\cite{Wootters}%
, and then by Bartell\cite{Bartell}. Some relevant inequalities to quantify
the interferometric duality can be found in a number of other publications%
\cite{Englert,Greenberger,Mandel,Jaeger1,Englert1}. Among them, Englert\cite
{Englert} presented definitions of the predictability $P$ and the
distinguishability $D$ to quantify how much WW information is stored in the
marker, and derived an inequality $D^{2}+V^{2}\leq 1$ at the intermediate
stage which puts a bound on $D$ when given a certain fringe visibility $V$.
Although the quantitative aspects of the interferometric complementarity
have been discussed by a number of theoretical papers, there are just a few
experimental studies, i.e., the neutron experiments\cite{Rauch,Summhammer},
the photon experiments\cite{Mittelstaedt,Schwindt} and the atom
interferometer\cite{Durr}. Recently, a complementarity experiment with an
interferometer at the quantum-classical boundary\cite{Bertet} was also
testified.

In this paper, we experimentally investigate the interferometric
complementarity of the ensemble-averaged spin states of one of two kinds of
nuclei in NMR sample molecules for the intermediate situations. We follow
our approach detailed in Ref. \cite{Zhu} but use two non-orthogonal spin
states of another nuclei in the sample molecules as the path markers. By
entangling the observed spin with the marker one, interference is destroyed
because it is in principle possible to determine the states the observed
spin possesses by performing a suitable measurement of the marker one\cite
{Englert}. However, in this paper, an alternative geometric strategy of
measuring $D$ is given and the relationship between $D$ and the entanglement
of the spin states is clarified. And finally the duality relation $%
D^{2}+V^{2}=1$ for various values of $D$ and $V$ is testified.

\section{Scheme and definition}

Our experimental scheme can be illustrated by a Mach-Zehnder interferometer
(shown in Fig. 1), a modified version of the two-slit experiment. The
observed and marker quantum objects, represented by B and A respectively,
compose a bipartite quantum system BA. Suppose the input state of BA to be $%
|\psi _{0}\rangle =\left| 0\right\rangle _{B}\left| 0\right\rangle
_{A}\equiv \left| 00\right\rangle ,$ with $\left| 0\right\rangle $ being one
of two orthonormal basis $\left| 0\right\rangle $ and $\left| 1\right\rangle 
$ of B and A. Firstly, a beam splitter (BS) splits $\left| 0\right\rangle
_{B}$ into $\frac{1}{\sqrt{2}}\left( \left| 0\right\rangle _{B}+\left|
1\right\rangle _{B}\right) ,$ meaning that the observed system B evolves
along two paths $\left| 0\right\rangle _{B}$ and $\left| 1\right\rangle _{B}$
simultaneously with equal probabilities. In the meantime, path markers (PM)
label the different paths $\left| 0\right\rangle _{B}$ and $\left|
1\right\rangle _{B}$ with the marker states $\left| m_{+}\right\rangle _{A}$
and $\left| m_{-}\right\rangle _{A}$ correspondingly. The joint action of
the BS and PM denoted by operation $U_{1}$, thus transforms $|\psi
_{0}\rangle $ into 
\begin{equation}
|\psi _{1}\rangle =\frac{1}{\sqrt{2}}(|0\rangle _{B}|m_{+}\rangle
_{A}+|1\rangle _{B}|m_{-}\rangle _{A}).
\end{equation}
Secondly, phase shifters (PS)\ add a relative phase difference between the
two paths, which are then combined into the output state $|\psi _{2}\rangle $
by a beam merge (BM). The joint action of the PS and BM, which is applied on
B solely, is accomplished by a unitary operation 
\begin{equation}
U_{2}=\frac{1}{\sqrt{2}}\left( 
\begin{array}{cc}
1 & e^{i\phi } \\ 
-e^{-i\phi } & 1
\end{array}
\right) .
\end{equation}
And the output state $|\psi _{2}\rangle =U_{2}|\psi _{1}\rangle $ could be
read as 
\begin{equation}
|\psi _{2}\rangle =\frac{1}{2}\left[ |0\rangle _{B}\left( |m_{+}\rangle
_{A}+e^{i\phi }|m_{-}\rangle _{A}\right) +|1\rangle _{B}\left( |m_{-}\rangle
_{A}-e^{-i\phi }|m_{+}\rangle _{A}\right) \right] .
\end{equation}

Finally, measuring the population, $I,$ of B in the state $|0\rangle _{B}$
and $|1\rangle _{B}$ gives 
\begin{equation}
I(\phi )=\frac{1}{2}(1\pm 
\mathop{\rm Re}%
(_{A}\langle m_{+}|m_{-}\rangle _{A})e^{i\phi }).
\end{equation}
where ``$\pm $'' correspond to the population in $|0\rangle _{B}$ and $%
|1\rangle _{B},$ respectively. Repeating the measurements at different $\phi 
$ might produce population fringes. Suppose the marker states $|m_{\pm
}\rangle _{A}=cos\varphi _{\pm }|0\rangle _{A}+sin\varphi _{\pm }|1\rangle
_{A},$ and from the usual definition of the fringe visibility $%
V=(I_{max}-I_{min})/\left( I_{max}+I_{min}\right) $ and Eq. (4) one gets

\begin{equation}
V=\left| _{A}\langle m_{+}|m_{-}\rangle _{A}\right| =\left| \cos \varphi
\right| ,
\end{equation}
where $\varphi =\varphi _{-}-\varphi _{+}.$

Englert\cite{Englert} proposed a quantitative measure for $D$ by introducing
a physical quantity $L_{W}$---the ``likelihood for guessing the right way'',
which depends on the choice of an observable $W,$ 
\begin{equation}
L_{W}=\sum\limits_{i}\max \left\{ p\left( W_{i},\left| 0\right\rangle
_{B}\right) ,p\left( W_{i},\left| 1\right\rangle _{B}\right) \right\} ,
\end{equation}
where $p\left( W_{i},\left| 0\right\rangle _{B}\right) $ and $p\left(
W_{i},\left| 1\right\rangle _{B}\right) $ denote the joint probabilities
that the eigenvalue $W_{i}$ of $W$ is found and the observed object takes
path $\left| 0\right\rangle _{B}$ or $\left| 1\right\rangle _{B}$. For
example, for the state of Eq. (1), an optimal observable $W_{opt}$ can be
found to maximize $L_{W}=\left( 1+\left| \sin \varphi \right| \right) /2$ in
the experiments\cite{Durr} and by the definition of the distinguishability $D
$ of paths $D=-1+2\max_{W}\left\{ L_{W}\right\} $\cite{Englert}, one gets 
\begin{equation}
D\left( \varphi \right) =\left| \sin \varphi \right| .
\end{equation}
Here, we present an expression for $D$ in an intuitively geometric way. To
this end, one projects the marker states $|m_{\pm }\rangle _{A}$ into an
appropriate orthonormal basis $\{|\beta _{+}\rangle _{A},|\beta _{-}\rangle
_{A}\},$%
\begin{equation}
\begin{array}{l}
|m_{+}\rangle _{A}=\gamma _{+}|\beta _{+}\rangle _{A}+\gamma _{-}|\beta
_{-}\rangle _{A}, \\ 
|m_{-}\rangle _{A}=\delta _{+}|\beta _{+}\rangle _{A}+\delta _{-}|\beta
_{-}\rangle _{A},
\end{array}
\end{equation}
where $\left| \gamma _{+}\right| ^{2}+\left| \gamma _{-}\right| ^{2}=\left|
\delta _{+}\right| ^{2}+\left| \delta _{-}\right| ^{2}=1.$ In the two-path
case the criterion of choosing $\{|\beta _{+}\rangle _{A},|\beta _{-}\rangle
_{A}\}$ is to make the difference of probabilities of measuring the two
states $|m_{+}\rangle _{A}$ and $|m_{-}\rangle _{A}$ on the basis $|\beta
_{+}\rangle _{A}$ to be equal to that while measuring $|m_{+}\rangle _{A}$
and $|m_{-}\rangle _{A}$ on $|\beta _{-}\rangle _{A}$. These probability
differences are then defined as the distinguishability 
\begin{equation}
D=\left| \left| \gamma _{+}\right| ^{2}-\left| \delta _{+}\right|
^{2}\right| =\left| \left| \delta _{-}\right| ^{2}-\left| \gamma _{-}\right|
^{2}\right| .
\end{equation}
The basis $\{|\beta _{+}\rangle _{A},|\beta _{-}\rangle _{A}\}$ can be
rewritten into in the computational basis:
\begin{equation}
\begin{array}{l}
|\beta _{+}\rangle _{A}=cos\theta |0\rangle _{A}+sin\theta |1\rangle _{A},
\\ 
|\beta _{-}\rangle _{A}=sin\theta |0\rangle _{A}-cos\theta |1\rangle _{A},
\end{array}
\end{equation}
where $\theta $ is the angle of the state vector $|\beta _{+}\rangle _{A}$
with respect to the basis $|0\rangle _{A}$. In order to satisfy Eq. (9),
from Fig. 2 and by the geometric knowledge $\theta =\frac{\varphi
_{+}+\varphi _{-}}{2}-\frac{\pi }{4}$ must be held, which yields

\begin{equation}
\begin{tabular}{l}
$\gamma _{+}=\delta _{-}=\cos (\frac{\pi }{4}-\frac{\varphi }{2}),$ \\ 
$\gamma _{-}=\delta _{+}=\sin (\frac{\pi }{4}-\frac{\varphi }{2}).$%
\end{tabular}
\end{equation}
Here, $\varphi =\varphi _{-}-\varphi _{+}$ is the angle between the two
marker state vectors in the Hilbert space. So from Eqs. (9) and (11), the
distinguishability is equally given by Eq. (7). It can also be seen that the
desired basis $\{|\beta _{+}\rangle _{A},|\beta _{-}\rangle _{A}\}$ deduced
by our geometric strategy is just the eigenvectors of the optimal observable 
$W_{opt}$\cite{Durr}.

These expressions for $V$ and $D$ are consistent with those in Ref. \cite
{Durr} and lead to the duality relation 
\begin{equation}
D^{2}\left( \varphi \right) +V^{2}\left( \varphi \right) =1.
\end{equation}

Eqs. (5) and (7) reveal the sinusoidal and cosinusoidal behaviors of $D$ and 
$V$, respectively, on the angle $\varphi $ between $|m_{+}\rangle _{A}$ and $%
|m_{-}\rangle _{A}$ in the Hilbert space $H_{A}$. $D$ and $V$, therefore,
are determined by the feature of $|m_{+}\rangle _{A}$ and $|m_{-}\rangle
_{A},$ especially by the value of $\varphi .$ However, for any value of $%
\varphi $ the duality relation (13) holds when two evolution paths, $%
|0\rangle _{B}$ and $|1\rangle _{B},$ are labeled by quantum pure state $%
|m_{+}\rangle _{A}$ and $|m_{-}\rangle _{A}.$ Generally Eq. (12) should be
replaced by $D^{2}+V^{2}\leq 1$\cite{Englert,Jaeger1,Englert1}.

As WW information of the observed system B is stored in the states of the
marker system A through the interaction and correlation of A and B, the
distinguishability of the B's paths depends on the feature of the marker
states, or more exactly, the correlation property of the combined system AB.
It would be natural to examine the relationship between the entanglement of
the system AB and the distinguishability. For a bipartite pure state the
entanglement $E$ can be denoted by the von Neumann entropy $S$\cite{Bennett}%
, $S=S\left( \rho ^{(A)}\right) =S\left( \rho ^{(B)}\right) ,$ with $S\left(
\rho ^{A(B)}\right) =-Tr(\rho ^{A(B)}\log _{2}\rho ^{A(B)})$ and $\rho
^{A(B)}=Tr_{A\left( B\right) }\left( \rho _{AB}\right) $ for each subsystem.
The entanglement $E$ for the pure state $|\psi _{1}\rangle $ shown in Eq.
(1) is then derived as

\begin{equation}
E\left( \varphi \right) =-\frac{1-\cos \varphi }{2}\log _{2}\left( \frac{%
1-\cos \varphi }{2}\right) -\frac{1+\cos \varphi }{2}\log _{2}\left( \frac{%
1+\cos \varphi }{2}\right) .
\end{equation}
It can be obtained from Eq. (13) that $E=0$ for $\varphi =k\pi $ and $E=1$
for $\varphi =\left( 2k+1\right) \pi /2$ with $k=0,1,2,\cdots ,$ which
correspond to $D=0$ and $1,$ respectively. A detailed quantitative analysis
of $E$ will be given later (see Fig. 3 below).

\section{Experimental procedure and results}

The scheme stated above was implemented by liquid-state NMR spectroscopy
with a two-spin sample of carbon-13 labeled chloroform $^{13}$CHCl$_{3}$
(Cambridge Isotope Laboratories, Inc.). We made use of the hydrogen nucleus (%
$^{1}$H) as the marker spin A and the carbon nuclei ($^{13}$C) as the
observed spin B in the experiments. Spectra were recorded on a BrukerARX500
spectrometer with a probe tuned at 125.77MHz for $^{13}$C and at 500.13MHz
for $^{1}$H . The spin-spin coupling constant $J$ between $^{13}$C and $^{1}$%
H is 214.95 Hz. The relaxation times were measured to be $T_{1}=4.8$ $sec$
and $T_{2}=3.3$ $sec$ for the proton, and $T_{1}=17.2$ $sec$ and $T_{2}=0.35$
$sec$ for carbon nuclei.

At first, we prepared the quantum ensemble in an effective pure state $\rho
_{0}$ from the thermal equilibrium by line-selective pulses with appropriate
frequencies and rotation angles and a magnetic gradient pulse\cite{Peng}. $%
\rho _{0}$ has the same properties and NMR experimental results as the pure
state $|\psi _{0}\rangle =|00\rangle $. Then we transferred $\rho _{0}$ to
another state $\rho _{1}$ equivalent to the state $|\psi _{1}\rangle $ shown
in Eq. (1) for accomplishing the BS and PM actions by applying a Hadamard
transformation $H_{B}=\frac{1}{\sqrt{2}}\left( 
\begin{array}{rr}
1 & 1 \\ 
1 & -1
\end{array}
\right) $ on spin B and two unitary transformations 
\begin{equation}
\begin{array}{l}
P_{1}=\exp (-iE_{+}^{A}\sigma _{y}^{B}\varphi _{+}), \\ 
P_{2}=\exp (-iE_{-}^{A}\sigma _{y}^{B}\varphi _{-}),
\end{array}
\end{equation}
where $\sigma _{\eta }^{i}(\eta =x,y,z)$ are Pauli matrices of the spin {\it %
i}, $E_{\pm }^{i}=\frac{1}{2}(1_{2}\pm \sigma _{z}^{i})$ and $1_{2}$ is the $%
2\times 2$ unit matrix. These operations were implemented by the NMR pulse
sequence $Y_{A}(\varphi _{+}+\varphi _{-})X_{A}(\frac{\pi }{2}%
)J_{AB}(\varphi _{-}-\varphi _{+})X_{A}(-\frac{\pi }{2})X_{B}(\pi )Y_{B}(%
\frac{\pi }{2})$ to be read from left to right, where $Y_{A}(\varphi
_{+}+\varphi _{-})$ denotes an $\varphi _{+}+\varphi _{-}$ rotation about $%
\hat{y}$ axis on spin $A$ and so forth, and $J_{AB}(\varphi _{-}-\varphi
_{+})$ represents a time evolution of $(\varphi _{-}-\varphi _{+})/\pi J_{AB}
$ under the scalar coupling between spins $A$ and $B$. Finally, the PS and
BM operations were achieved by the transformation $U_{2}$, which was
realized by the NMR pulse sequence $X_{B}\left( -\theta _{1}\right)
Y_{B}\left( \theta _{2}\right) X_{B}\left( -\theta _{1}\right) $ with $%
\theta _{1}=\tan ^{-1}(-\sin \phi )$, and $\theta _{2}=2sin^{-1}(-cos\phi /%
\sqrt{2})$.

In our experiments, two sets of experiments for a given value of $\varphi
=\varphi _{-}-\varphi _{+}$ were performed to measure the fringe visibility $%
V$ and the distinguishability $D.$ In the experiment of a quantitative
measure for $D$, whether it is defined by the geometric way or the maximum
likelihood estimation, the joint probabilities $p\left( |\beta _{\pm
}\rangle _{A},\left| 0\right\rangle _{B}\right) $ and $p\left( |\beta _{\pm
}\rangle _{A},\left| 1\right\rangle _{B}\right) $ must firstly be measured.
We performed the joint measurements by a two-part procedure inspired by
Brassard et al.\cite{Brassard}. Part one of the procedure is to rotate from
the basis $\{|0\rangle _{B}|\beta _{+}\rangle _{A},|0\rangle _{B}|\beta
_{-}\rangle _{A},|1\rangle _{B}|\beta _{+}\rangle _{A},|1\rangle _{B}|\beta
_{-}\rangle _{A}\}$ into the computational basis $\{|00\rangle ,|01\rangle
,|10\rangle ,|11\rangle \}$ (omitting the subscripts A and B), which was
realized by the unitary operation 
\begin{equation}
R_{B}=\left( 
\begin{array}{ll}
\cos \alpha  & -\sin \alpha  \\ 
\sin \alpha  & \cos \alpha 
\end{array}
\right) 
\end{equation}
where $\alpha =\frac{\pi }{4}-\frac{\varphi _{+}+\varphi _{-}}{2}$,
corresponding to the NMR pulse $Y_{B}(2\alpha )$. Part two of the procedure
is to perform a projective measurement in the computational basis which
could be mimiced by a magnetic gradient pulse along $z$-axis\cite
{Teklemariam}. Accordingly, , the joint probabilities $p\left( |\beta _{\pm
}\rangle _{A},\left| 0\right\rangle _{B}\right) $ and $p\left( |\beta _{\pm
}\rangle _{A},\left| 1\right\rangle _{B}\right) $ were obtained with
reconstructing the diagonal elements of the deviation density matrix by
quantum state tomography\cite{Chuang}. The results are shown in Fig. 3. In
our geometric strategy, it can be obtained from Eqs. (1) and (8) that, the
information of $\gamma _{+},$ $\gamma _{-}$ or $\delta _{+},$ $\delta _{-}$
are determined by the population probabilities, i.e., $\left| \gamma
_{+}\right| ^{2}=2p(|0\rangle _{B}|\beta _{+}\rangle _{A}),\left| \gamma
_{-}\right| ^{2}=2p(|0\rangle _{B}|\beta _{-}\rangle _{A})$ and $\left|
\delta _{+}\right| ^{2}=2p(|1\rangle _{B}|\beta _{+}\rangle _{A})=,\left|
\delta _{-}\right| ^{2}=2p(|1\rangle _{B}|\beta _{-}\rangle _{A})$. Finally,
we used Eq. (9) and took the average value of $\left( \left| \left| \gamma
_{+}\right| ^{2}-\left| \delta _{+}\right| ^{2}\right| +\left| \left| \delta
_{-}\right| ^{2}-\left| \gamma _{-}\right| ^{2}\right| \right) /2$ to give
data points of $D$ which shown in Fig. 4. On the other hand, utilizing data
points of Fig. 3, we achieved the experimental values of the likelihood $%
L_{W}$ from Eq. (6) and obtained the $D$ measure with the maximum likelihood
estimation strategy, which is the same outcomes as that in our geometric
strategy. Therefore, the intuitively geometric strategy gives the equally
effective measure of the distinguishability $D$.

For measuring $V$, we repeatedly applied the NMR pulse sequence $X_{B}\left(
-\theta _{1}\right) Y_{B}\left( \theta _{2}\right) X_{B}\left( -\theta
_{1}\right) $ that represents the $U_{2}\left( \phi \right) $ operation for
various values of $\phi $ and detected the population of B in the state $%
\rho _{2}$ equivalent to the output state $\left| \psi _{2}\right\rangle .$
A set of appropriate values $\theta _{1}$ and $\theta _{2}$ were chosen to
vary the values for $\phi $ from $0$ to $2\pi .$ Using the same reading-out
pulses and tomography method as in the measurement of $D$, we reconstructed
the populations of B for various values of $\phi .$ The variation of the
normalized populations versus $\phi $ showed a desirable interference
fringe, from which the value of $V$ was extracted. Care should be exercised
in processing the spectra data of the different experimental runs in order
to get the normalized populations of the deviation density matrix.

The objective of the present paper is to study the interferometric
complementarity in the intermediate regime with two non-orthogonal marker
states, so the experimental procedure mentioned above was repeated for
different $\varphi .$ Without loss of generality, we assumed $\varphi _{+}=%
\frac{\pi }{2}$ and changed the $\varphi $ values from $0$ to $5\pi /4$ by
varying the $\varphi _{-}$ value with the increment of $\pi /16.$ The
measured values of $V\left( \varphi \right) $ and $D\left( \varphi \right) $
in two sets of independent experiments were plotted in Fig. 3, along with
the theoretical curves of $V\left( \varphi \right) ,$ $D\left( \varphi
\right) $ and $E\left( \varphi \right) .$ The experimental data and
theoretical curve for $D^{2}\left( \varphi \right) +V^{2}\left( \varphi
\right) $ were depicted in Fig. 4.

From Figs. 3 and 4 some remarks can be made as follows.

1) For $\varphi =k\pi ,(k=0,1,2,\cdots )$ which means $\left| _{A}\langle
m_{+}|m_{-}\rangle _{A}\right| =1,$ two marker states are identical
(differing with an irrelevant phase factor possible), and the state of the
system AB\ is completely unentangled $(E=0)$. In this case no WW information
of system B is stored in system A so that two evolution paths of B is
indistinguishable $(D=0)$ and perfect fringe visibility is observed $(V=1)$.
For $\varphi =(2k+1)\pi /2,$ i.e., $\left| _{A}\langle m_{+}|m_{-}\rangle
_{A}\right| =0,$ the marker states are orthogonal, and the state of the
system AB\ is completely entangled $(E=1)$. This leads to full WW
information $(D=1)$ and no interference fringes $(V=0).$ These two extremes
are exactly the same examples that we have studied in Ref.\cite{Zhu} with a
NMR bulk ensemble by population measurements.

2) When $\varphi $ equals other values than $k\pi $ and $(2k+1)\pi /2,$
which corresponds to $0<\left| _{A}\langle m_{+}|m_{-}\rangle _{A}\right|
<1, $ the marker states are partially orthogonal, and the state of the AB\
system is partially entangled $(0<E<1)$. In this intermediate situations
partial fringe visibility $\left( 0<V<1\right) $ and partial WW information $%
\left( 0<D<1\right) $ are resulted. Nevertheless, the interferometric
duality still holds as in the extreme cases.

3) In the whole range of $\varphi ,$ $E$ varies synchronously with $D$. The
reason is that the increase of $E$ means more correlation between system B
and A and more WW information of B stored in A, so $D$ rises, and vice
versa. On the contrary, the variation trend of E versus $\varphi $ is
opposite to that of $V$ versus $\varphi .$ As the function of E versus $%
\varphi $ has a complicated form there is no similar relation between $E$
and $V$ to the duality of $D^{2}+V^{2}=1$.

4) The measured values of $V$, $D$ and thus the derived values of $%
D^{2}+V^{2}$ are fairly in agreement with the theoretical expectation. The
discrepancies between the experimental and theoretical values of $V$, $D$
and $D^{2}+V^{2}$ in some data points, estimated to be less than $\pm 10\%,$
are due to the inhomogeneity of the RF field and static magnetic field,
imperfect calibration of RF pulses, and signal decaying during the
experiments.

\section{Conclusion}

In conclusion, we have experimentally tested the interferometric
complementarity in a spin ensemble with NMR techniques. In addition to two
extremes, the intermediate cases that the fringe visibility $V$ reduces due
to the increase of the storage of WW information are emphasized. The
measured data of $D$ and $V$ in our NMR experiments are in consistent with
the duality relation. In particular, the close link among $D$, $V$ and the
entanglement of the composite system consisting of the observed and marker
states is explicitly revealed and explained. Though the experiment was not
strictly limited in the one-photon-at-a-time fact, it was performed on a
quantum ensemble whose dynamical evolution is still quantum mechanical.
Therefore, our experiment provides a test of the duality relation in the
intermediate situations.

\begin{center}
{\bf ACKNOWLEDGEMENTS}
\end{center}

This work was supported by the National Natural Science Foundation of China
(Grant NO. 1990413). X. Peng thanks Xiaodong Yang, Hanzeng Yuan, and Xu
Zhang for help in the course of experiments.

\begin{center}
{\large Figure Captions}
\end{center}

Fig. 1 Schematic diagram of a two-way interferometer. The input, say $%
|0\rangle _{B}|0\rangle _{A},$ is split to two ways by the beam splitter
(BS), then labeled by the path markers (PM), phase shifted by the phase
shifters (PS)\ and finally recombined into the output by the beam merger
(BM).

Fig. 2 The state vectors in Hilbert space for defining $D$. $\{|0\rangle
,|1\rangle \}$ represents the standard orthonormal basis. Two marker states $%
\left\{ |m_{\pm }\rangle \right\} $ and another orthonormal basis $\left\{
|\beta _{\pm }\rangle \right\} $ are determined by the angle $\varphi
_{+},\varphi _{-}$ and $\theta ,\frac{\pi }{2}+\theta ,$ respectively. The
dashed line denotes the angle bisector between the two states $\left\{
|m_{\pm }\rangle \right\} $. From the map, one can get the relation $\theta =%
\frac{\varphi _{+}+\varphi _{-}}{2}-\frac{\pi }{4}.$

Fig. 3 Normalized populations versus the angle $\varphi $, between two
marker states in the experiments to measure $D$. Data points $+,\bigcirc ,*$
and $\boxdot $ denote the joint probabilities $p(|\beta _{+}\rangle
_{A},|0\rangle _{B}),$ $p(|\beta _{-}\rangle _{A},|0\rangle _{B}),$ $%
p(|\beta _{+}\rangle _{A},|1\rangle _{B})$ and $p(|\beta _{-}\rangle
_{A},|1\rangle _{B})$, respectively. Theoretical curves expressed by $%
p(|\beta _{+}\rangle _{A},|0\rangle _{B})=p(|\beta _{-}\rangle
_{A},|1\rangle _{B})=\left| \cos (\frac{\pi }{4}-\frac{\varphi }{2})\right|
^{2}/2$ and $p(|\beta _{-}\rangle _{A},|0\rangle _{B})=p(|\beta _{+}\rangle
_{A},|1\rangle _{B})=\left| \sin (\frac{\pi }{4}-\frac{\varphi }{2})\right|
^{2}/2$ are depicted with the solid line and the dashdotted line,
respectively.

Fig. 4 Visibility $V$ (denoted by $\bigcirc $) and distinguishability $D$
(denoted by $*$) as a function of $\varphi $. The solid lines are the
theoretical expectations of $V$ and $D$ and the dashed line denotes $E$
expressed by Eq. (11).

Fig. 5 Experimental test of the duality relation based on the data from Fig.
3. $D^{2}+V^{2}$ is plotted as a function of $\varphi $. The solid line
represents the theoretical expectation.


\begin{references}
\bibitem{Bohr}  N. Bohr, 1928 Naturwissenschaften {\bf 16} 245; 1928 Nature
(London) {\bf 121} 580.

\bibitem{Englert}  B. -G. Englert, 1996 Phys. Rev. Lett. {\bf 77} 2154.

\bibitem{Feynman}  R. P. Feynman, R. B. Leifhton, and M. Sands, the Feynamn
Lectures of Physics, Vol. III. Quantum Mechanics, Addison -Wesley, Reading
(1965).

\bibitem{Taylor}  G. I. Taylor, 1909 Proc. Camb. Phil. Soc. {\bf 15} 114.

\bibitem{Mollenstedt}  G. M$\ddot{o}$llenstedt and C. J$\ddot{o}$nsson, 1959
Z. Phys. {\bf 155} 472; A. Tonomura, J. Endo, T. Matsuda, and T. Kawasaki,
1989 Am. J. Phys. {\bf 57} 117.

\bibitem{Zeilinger}  A. Zeilinger, R. G$\ddot{a}$hler, C. G. Shull, W.
treimer, and W. Mampe, 1988 Rev. Mod. Phys. {\bf 60} 1067.

\bibitem{Carnal}  O. Carnal and J. Mlynek, 1991 Phys. Rev. Lett. {\bf 66}
2689.

\bibitem{Zhu}  X. Zhu, X. Fang, X. Peng, M. Feng, K. Gao and F. Du, 2001 J.
Phys. B {\bf 34} 4349.

\bibitem{recoiling}  N. Bohr, in Albert Einstein: Philosopher Scientist(ed.
P. A. Schilpp) 200-241 (Library of Living Philosophers, Evanston, 1949);
reprinted in Quantum Theory and Measurement (eds J. A. Wheeler and W. H.
Zurek) 9-49 (Princeton Univ. Press, Princeton, 1983).

\bibitem{Scully}  M. O. Scully, B. -G. Englert and H. Walther, 1991 Nature 
{\bf 351} 111.

\bibitem{Wootters}  W. K. Wootters and W. H. Zurek, 1979 Phys. Rev. D {\bf 19%
} 473.

\bibitem{Bartell}  L. S. Bartell, 1980 Phys. Rev. D {\bf 21} 1698.

\bibitem{Greenberger}  D. M. Greenberger and A. Yasin, 1988 Phys. Lett. A 
{\bf 128} 391.

\bibitem{Mandel}  L. Mandel, 1991 Opt. Lett. {\bf 16} 1882.

\bibitem{Jaeger1}  G. Jaeger, A. Shimony and L. Vaidman, 1995 Phys. Rev. A 
{\bf 51} 54.

\bibitem{Englert1}  B. -G. Englert, J. A. Bergou, 2000 Opt. Comm. {\bf 179}
337.

\bibitem{Rauch}  H. Rauch and J. Summhammer, 1984 Phys. Lett. A {\bf 104} 44.

\bibitem{Summhammer}  J. Summhammer, H. Rauch and D. Tuppinger, 1987 Phys.
Rev. A {\bf 36} 4447.

\bibitem{Mittelstaedt}  P. Mittelstaedt, A. Prieur and R. Schieder, 1987
Found. Phys. {\bf 17} 891.

\bibitem{Schwindt}  P. D. D. Schwindt, P. G. Kwiat and B. -G. Englert, 1999
Phys. Rev. A {\bf 60} 4285.

\bibitem{Durr}  S. D$\ddot{u}$rr, T. Nonn and G. Rempe, 1998 Phys. Rev.
Lett. {\bf 81} 5705.

\bibitem{Bertet}  P. Bertet, S. Osnaghl, A. Rauschenbeutel, G. Nogues, A.
Auffeves, M. Brune, J. M. Ralmond and S. Haroche, 2001 Nature {\bf 411} 166.

\bibitem{Scully1}  M. O. Scully and M. S. Zubairy, Quantum optics (Cambridge
Univ. Press, Cambridge, UK, 1997).

\bibitem{Bennett}  C. H. Bennett,H. J. Bernstein S. Popescu B. Schumacher,
1996 Phys. Rev. A {\bf 53} 2046.

\bibitem{Peng}  X. Peng, X. Zhu, X. Fang, M. Feng, K. Gao, X. Yang and M.
Liu, 2001 Chem. Phys. Lett. {\bf 340} 509.

\bibitem{Brassard}  G. Brassard, S. Braunstein and R. Cleve, 1998 Phys. D 
{\bf 120} 43.

\bibitem{Teklemariam}  G. Teklemariam, E. M. Fortunato, M. A. Pravia, T. F.
Havel and D. G. Cory, 2001 Phys. Rev. Lett. 86, 5845.

\bibitem{Chuang}  I. L. Chuang, N. Gershenfeld, M. Kubinec and D. Leung,
1998 Proc. Roy. Soc. Lond A {\bf 454} 447.\newpage 
\end{references}
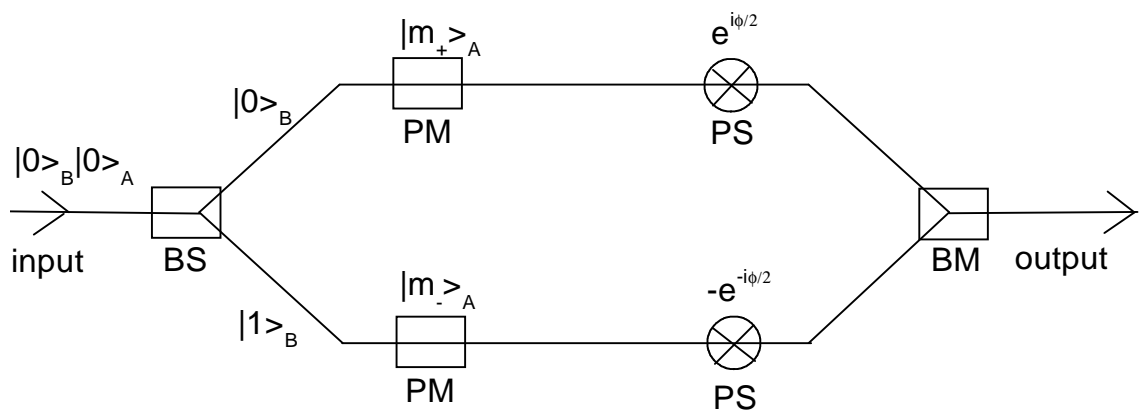
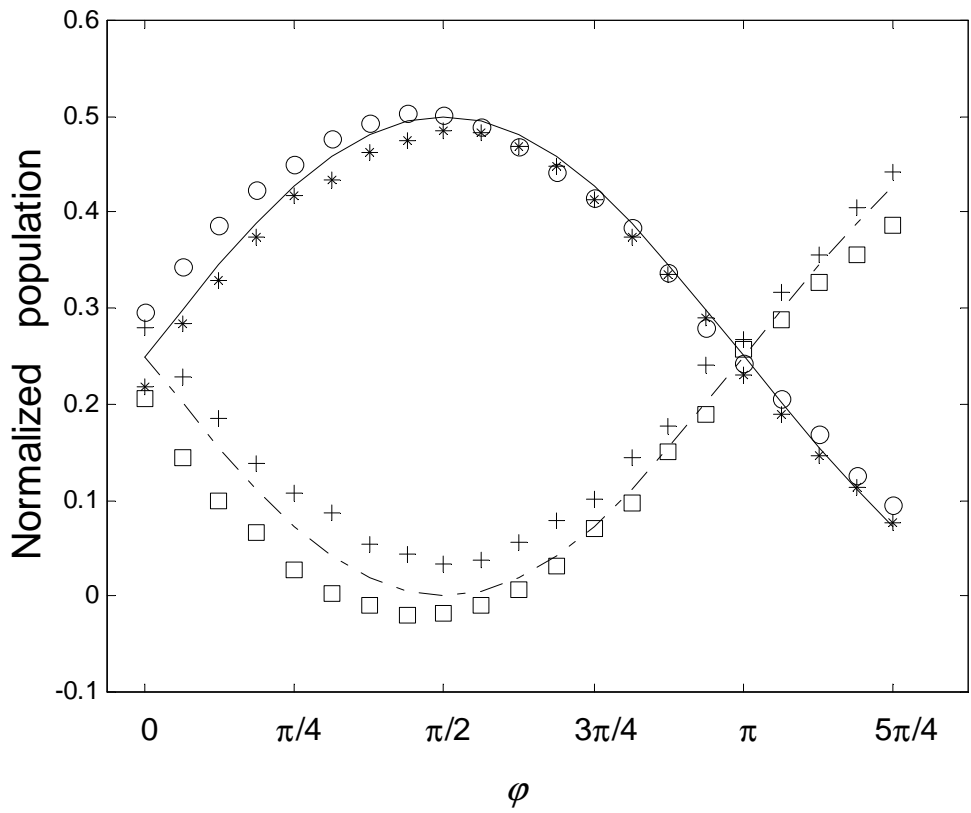
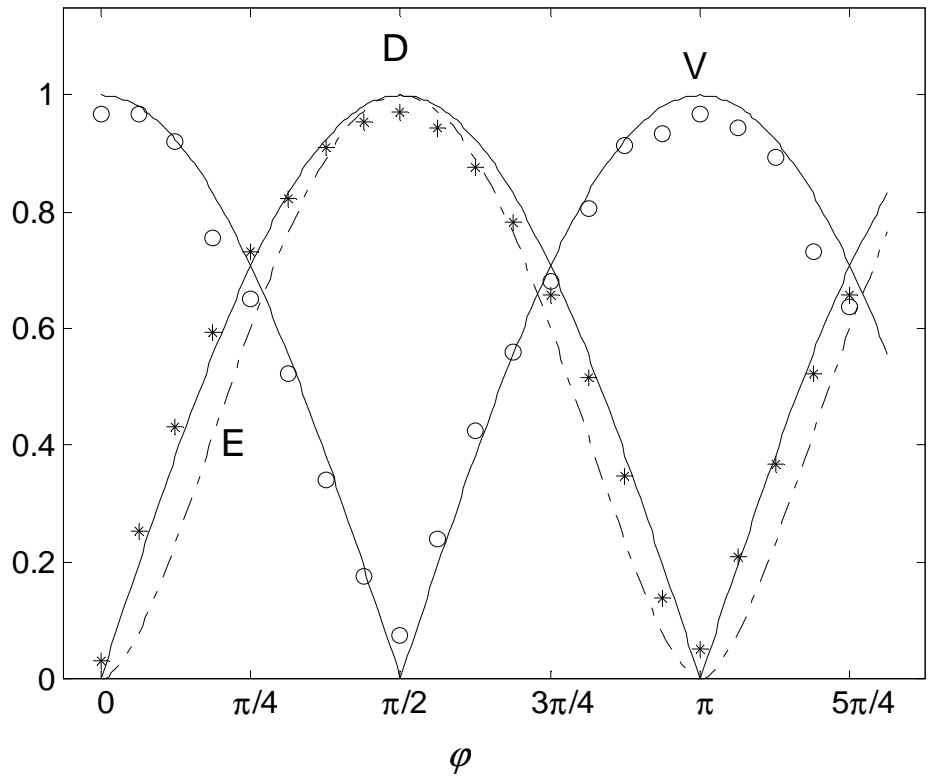
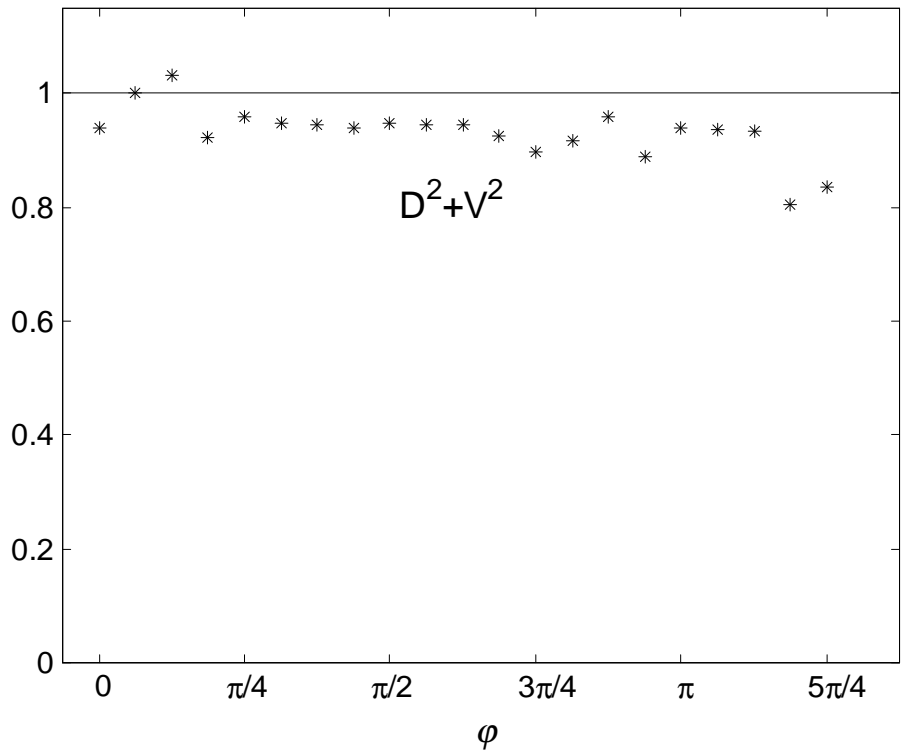
\end{document}